\documentclass[letter]{jpsj2} 
\usepackage{color}

\title{Electric-dipole active two-magnon excitation in {\textit{ab}} spiral spin phase of a ferroelectric magnet Gd$_{\textbf{0.7}}$Tb$_{\textbf{0.3}}$MnO$_{\textbf 3}$}

\author{Noriaki \textsc{Kida}$^{1}$\thanks{E-mail: n-kida@erato-mf.t.u-tokyo.ac.jp}, Yuichi \textsc{Yamasaki}$^{2}$, Ryo \textsc{Shimano}$^{1,3}$, Taka-hisa \textsc{Arima}$^{4}$, and Yoshinori \textsc{Tokura}$^{1,2,5}$}

\inst{$^{1}$Multiferroics Project (MF), ERATO, Japan Science and Technology Agency (JST), c/o Department of Applied Physics, The University of Tokyo, 7-3-1 Hongo, Bunkyo-ku, Tokyo 113-8656, Japan \\
$^{2}$Department of Applied Physics, The University of Tokyo, 7-3-1 Hongo, Bunkyo-ku, Tokyo 113-8656, Japan \\
$^{3}$Department of Physics, The University of Tokyo, 7-3-1 Hongo, Bunkyo-ku, Tokyo 113-0033, Japan \\
$^{4}$Institute of Multidisciplinary Research for Advanced Materials, Tohoku University, 2-1-1 Katahira, Aoba-ku, Sendai 980-8577, Japan \\
$^{5}$Cross-Correlated Materials Research Group (CMRG), ASI, RIKEN, 2-1 Hirosawa, Wako, 351-0198, Japan}

\abst{A broad continuum-like spin excitation (1--10 meV) with a peak structure around 2.4 meV has been observed in the ferroelectric $ab$ spiral spin phase of Gd$_{0.7}$Tb$_{0.3}$MnO$_3$ by using terahertz (THz) time-domain spectroscopy. Based on a complete set of light-polarization measurements, we identify the spin excitation active for the light $E$ vector only along the $a$-axis, which grows in intensity with lowering temperature even from above the magnetic ordering temperature but disappears upon the transition to the $A$-type antiferromagnetic phase. Such an electric-dipole active spin excitation as observed at THz frequencies can be ascribed to the two-magnon excitation in terms of the unique polarization selection rule in a variety of the magnetically ordered phases.}

\kword{electric-dipole active two-magnon excitation, electromagnon, spin excitation, magnon, multiferroics, terahertz time-domain spectroscopy}

\begin{document}
\maketitle

There has been continuous interest in revealing the low-energy electrodynamics in solids with electric and/or magnetic order. A typical example is the soft mode of the displacement type ferroelectrics, as manifested in the complex dielectric constant spectrum $\epsilon(\omega)$, as a response of the light electric field $E^\omega$. The spin excitation in (anti)ferromagnets driven by the magnetic field of light $H^\omega$, characterized by the complex magnetic permeability spectrum $\mu(\omega)$, is another typical example. Recently, the presence of the spin excitation driven by $E^\omega$ rather than $H^\omega$ was proposed to explain the optical spectra at terahertz (THz) frequencies of the ferroelectric magnet,\cite{MEeffectRev} orthorhombically distorted perovskite manganite, TbMnO$_3$.\cite{APimenov1} Such an electric-dipole active spin excitation, as theoretically anticipated to exist,\cite{VGBaryakhtar} is now referred to as $electromagnon$.\cite{APimenov1} In the magnetically driven ferroelectric phase of TbMnO$_3$,\cite{TKimura1,TKimura2} the spiral spin order has been experimentally confirmed,\cite{MKenzelmann} being consistent with the spin-current mechanism\cite{HKatsura1} or the inverse Dzyaloshinskii-Moriya effect;\cite{IASergienko,MMostovoy} the electric polarization $P_{\rm s}$ can be generated, as formulated as $P_{\rm s}\propto \sum_{<ij>} e_{ij}\times(S_i \times S_j)$, where $e_{ij}$ is the unit vector connecting to the neighboring spins, $S_i$ and $S_j$ (left panel of Fig. \ref{fig1}). Noticeably, $P_{\rm s}$ can be flopped from the $c$- to $a$-axis for TbMnO$_3$ and DyMnO$_3$ [see Fig. \ref{fig1}(a)] when the magnetic field $H$ exceeds the critical value $H_{\rm c}$.\cite{TKimura1,TKimura2} This can be regarded as $H$ control of the spiral spin plane from $bc$ to $ab$.

Up to date, the electromagnon or electric-dipole active spin excitation was argued to emerge in $\epsilon(\omega)$ around 2 meV in a family of $R$MnO$_3$ ($R=$ Tb, Dy, Gd, and Eu$_{1-x}$Y$_x$ with nominal composition $x$) by the THz or far-infrared optical spectroscopy.\cite{APimenov1,APimenov2,NKida,RValdesAguilar,APimenov3} As the origin of this excitation, the rotation mode of the spiral spin plane was theoretically considered on the basis of the spin-current mechanism.\cite{HKatsura2} Such a rotation mode should become active when $E^\omega$ is set perpendicular to the spiral spin plane. In fact, a remarkable spin excitation with a peak structure around 3 meV is observed in $\epsilon(\omega)$ for $E^\omega\parallel a$ in the $bc$ spiral spin phase of TbMnO$_3$ in zero $H$.\cite{APimenov1} In a recent inelastic neutron scattering study, one of low-energy branches of the magnon band was assigned to the $E^\omega\parallel a$ electromagnon around 2.5 meV at $k=0$, and hence to the rotation of the spiral spin plane.\cite{DSenff} However, this issue remains controversial by a recent observation that the spin excitation for $E^\omega\parallel a$ in DyMnO$_3$ well survives in the $H$-induced $ab$ spiral spin phase.\cite{NKida} Furthermore, the analogous absorption is observed also for $E^\omega\parallel a$ in Eu$_{1-x}$Y$_x$MnO$_3$ even though $P_{\rm s}$ in this system emerges along the $a$-axis in zero $H$.\cite{RValdesAguilar,APimenov3}

To settle the above ambiguous situation of the experimental results of the electric-dipole active spin excitation, we here investigate a complete set of light-polarization dependence of the optical spectra at THz frequencies for Gd$_{0.7}$Tb$_{0.3}$MnO$_3$, which hosts a variety of thermally induced phases, including the $ab$ spiral spin phase [see Fig. \ref{fig1}(b)]. We observed in this study that the spin excitation is strongly active only for the light $E$ vector along the $a$-axis and hence it cannot be ascribed to the rotation mode of the spiral spin plane. The electric-dipole active two-magnon process can qualitatively explain the observed characteristics of the spin excitations in a variety of thermally induced spin phases.

Since spin excitations may be driven by either $E^\omega$ or $H^\omega$ in a family of $R$MnO$_3$, it is indispensable to measure the complete set of optical spectra for the isolation of the components of $\epsilon$ and $\mu$. The phase diagram for Gd$_{0.7}$Tb$_{0.3}$MnO$_3$ in a plane of $T$ and $H$ is reproduced in Fig. \ref{fig1}(b).\cite{TGoto2} The paraelectric (PE) collinear spin order evolves along the $b$-axis below $T_{\rm N}^{\rm Mn}=42$ K, as in the similar manner to other $R$MnO$_3$.\cite{TKimura1,TKimura2} Below the ferroelectric (FE) transition temperature $T_{\rm c}=24$ K $(P_{\rm s} \parallel a)$, the $ab$ spiral spin order was recently identified by polarized neutron scattering experiments,\cite{YYamasaki} as schematically shown in the left panel of Fig. \ref{fig1}. Furthermore, the modulation wavenumber $q_b^{\rm Mn}$ of Mn spins along the $b$-axis was found to be 0.25, which is identical to $q_b^{\rm Mn}$ of the FE $ab$ spiral spin phase of TbMnO$_3$ induced by $H$.\cite{TArima} Accordingly, the $H$ induced $ab$ spiral spin phase of TbMnO$_3$ [Fig. \ref{fig1}(a)] smoothly connects to the thermally induced $ab$ spiral spin phase of Gd$_{0.7}$Tb$_{0.3}$MnO$_3$ [Fig. \ref{fig1}(b)]. Below $T_{\rm N}^{\rm WFM}=15$ K, $P_{\rm s}$ vanishes and the $A$-type antiferromagnetic (AFM) phase is formed with a weak ferromagnetic moment along the $c$-axis as a result of the spin canting. Therefore, Gd$_{0.7}$Tb$_{0.3}$MnO$_3$ can provide the ideal platform for further getting the insights into the nature of the spin excitation, especially in the $ab$ spiral spin phase, as realized in TbMnO$_3$ in high $H$.

We employed the THz time-domain spectroscopy in a transmission geometry. The direction of the light polarization was carefully set parallel to the crystallographic axis with use of wire grid polarizers.  We estimated the complex optical constants $\tilde{n}$ (=$\sqrt{\epsilon\mu}$) from transmission measurements for that the contribution of $\mu$ to the Fresnel coefficient was confirmed to be negligible in the case of the thick samples. Details of our experimental setup and the validity of adopting $\epsilon\mu(\omega)$ are described in detail in ref. 11. The single-crystalline samples were grown by a floating zone (FZ) method.\cite{TGoto2} Thin plates with wide $ac$, $ab$, and $bc$ faces were cut and polished to the thickness of 100--700 $\mu$m.

First, we overview the selection rule of the spin excitation based on the complete set of the light-polarization (both $E^\omega$ and $H^\omega$) measurements using available $ac$-, $ab$-, and $bc$-surface crystal plates. In Fig. \ref{fig2} we present the temperature $(T)$ variation of $\epsilon\mu$ spectra (closed circles) in $A$-type AFM ($T<15$ K), FE $ab$ spiral spin ($P_{\rm s}\parallel a$; 15 K $<T<$ 24 K), PE collinear spin (24 K $<T<$ 42 K), and PE paramagnetic (42 K $<T$) phases. The upper and lower panels show the real and imaginary parts of $\epsilon\mu$ spectra, respectively. There are spectral similarities in the observed light-polarization dependence, apart from in the $A$-type AFM phase, to those of DyMnO$_3$ in zero $H$.\cite{NKida} Especially, in the FE $ab$ spiral spin phase, the broad continuum-like absorption, that was assigned to the electric-dipole active spin excitation or electromagnon in the FE $bc$ spiral spin phase of $R$MnO$_3$,\cite{APimenov1,NKida} can be clearly seen in $Im[\epsilon\mu]$ spectra around 2.4 meV when $E^\omega$ was set parallel to the $a$-axis [Figs. \ref{fig2}(a) and \ref{fig2}(b)]. Accordingly, it yields the large optical anisotropy of $Re[\epsilon\mu]$ in the lowest energy region. Incidentally, there is a difference in the magnitude of $Im[\epsilon\mu]$ for $E^\omega \parallel a$ using $ac$- and $ab$-surface crystal plates. We carefully checked the reproducibility of $\epsilon\mu$ spectra by changing the thickness of the samples. Even though thin plates were cut from the nearly same region ($\sim$ 1 cm$^3$) of the FZ crystal, such a discrepancy may arise from a slight difference of $x$ in respective plates, as Gd$_{0.7}$Tb$_{0.3}$MnO$_3$ locates close to the bicritical point between competing FE and PE phases in a plane of $x-T$.\cite{TGoto2} However, discussions described below are not affected qualitatively by these discrepancies. In the following, we clarify the selection rule of the spin excitation in a variety of thermally induced phases.

In the $A$-type AFM phase at 10 K, a sharp peak structure can be seen around 2.2 meV in $Im[\epsilon\mu]$ spectrum for $E^\omega \parallel c$ and $H^\omega \parallel a$ with a clear dispersion in $Re[\epsilon\mu]$ spectrum [Fig. \ref{fig2}(d)]. This absorption can be ascribed to the spin excitation driven by $H^\omega \parallel a$, since the nearly same spectral signature can be also identified in $\epsilon\mu$ spectrum at 10 K for $E^\omega \parallel b$ and $H^\omega \parallel a$ [Fig. \ref{fig2}(e)]. We also observe the sharp peak structure at 10 K for $E^\omega \parallel b$ and $H^\omega \parallel c$ [Fig. \ref{fig2}(f)], which is also discerned in $\epsilon\mu$ spectrum at 11 K for $E^\omega \parallel a$ and $H^\omega \parallel c$ [Fig. \ref{fig2}(a)], clearly indicating that there is also spin excitation driven by $H^\omega \parallel c$. Below 15 K, the ferromagnetic moment ($\sim0.2$ $\mu_{\rm B}$/Mn-site) appears along the $c$-axis, activating the spin excitation for $H^\omega \parallel c$ [Fig. \ref{fig1}(a)]. These spin excitations can be interpreted as AFM resonances (AFMRs) of Mn ions, which are also seen in canted AFM phases of Eu$_{0.9}$Y$_{0.1}$MnO$_3$ for $H^\omega \parallel a$ (ref. 13) and La$_{1-x}$Sr$_x$MnO$_3$ $(x<0.1)$ for $H^\omega \parallel c$ (ref. 19). In fact, the $k=0$ magnon in the $A$-type AFM phase of LaMnO$_3$ locates around 3 meV.\cite{KHirota} Therefore, the $\epsilon\mu$ spectrum is considered to consist of a sharp resonance for $\mu$ and broad background absorption for $\epsilon$, which can be phenomenologically expressed with two Lorentz oscillators (solid lines in Fig. \ref{fig2}). In the $ab$ spiral spin phase above 15 K, such AFMRs still survive while the peak position shifts to the lower-energy. Further investigations on AFMRs are beyond the scope of this work, and in the following we focus on the spin excitation driven by $E^\omega$.

In the PE paramagnetic phase (e.g., at 201 K), no remarkable absorption is visible in $Im[\epsilon\mu]$ spectrum for $E^\omega \parallel a$ and $H^\omega\parallel c$, as shown in Fig. \ref{fig2}(a). With decreasing $T$, the broad continuum-like absorption, as ascribed to the spin excitation, is clearly identified in the PE collinear spin phase below $T_{\rm N}^{\rm Mn}\sim42$ K, as exemplified by $\epsilon\mu$ spectrum at 33 K. By comparing this with the $\epsilon\mu$ spectrum at the same $T$ (33 K) for $E^\omega \parallel a$ and $H^\omega\parallel b$ [Fig. \ref{fig2}(b)], this continuum-like intense band can be identified to be electric-dipole active only for $E^\omega\parallel a$. Apparently, it shows up even above $T_{\rm N}^{\rm Mn}=42$ K as a Debye-like absorption, the characteristic energy of which is estimated to be $\sim2.9$ meV, as indicated by solid lines in Figs. \ref{fig2}(a) and \ref{fig2}(b). In the vicinity of $T_{\rm c}=24$ K, the spin excitation for $E^\omega \parallel a$ grows in intensity and forms low- and high-lying peak structures around 2.4 meV and 8 meV, respectively, as seen in $\epsilon\mu$ spectrum at 23 K [Fig. \ref{fig2}(a)]. The position of the low-lying peak structure ($\sim2.4$ meV) is nearly identical to those observed for other $R$MnO$_3$.\cite{APimenov1,APimenov2,NKida,RValdesAguilar,APimenov3} At 17 K when the FE $ab$ spiral spin order is fully developed, the magnitude of $Im[\epsilon\mu]$ reaches the maximum $\sim4$. The corresponding optical conductivity is about 1.4 $\Omega^{-1}$cm$^{-1}$, being of the same order of other $R$MnO$_3$ (1--3 $\Omega^{-1}$cm$^{-1}$).\cite{APimenov1,APimenov2,NKida,RValdesAguilar,APimenov3} The $\epsilon\mu$ spectrum can be reproduced by two Lorentz oscillators for low- and high-lying peak structures (solid lines). It yields $Re[\epsilon(\omega\rightarrow0)]$ of $\sim24$, being comparable with the value of $Re[\epsilon]$ measured at 1 kHz $(\sim30)$. As further decreasing $T$ below 15 K at which the $A$-type AFM is realized, $Im[\epsilon\mu]$ is considerably suppressed. At 11 K, the finite peak structure is still visible [Fig. \ref{fig2}(a)], which is due to the presence of $\mu$ component driven by $H^\omega \parallel c$, as described before. This is firmly supported by the fact that there is no discernible peak absorption at 11 K for $E^\omega \parallel a$ and $H^\omega \parallel b$ [Fig. \ref{fig2}(b)].

To be more quantitative, we show in Fig. \ref{fig3} the $T$ dependence of the integrated spectral weight per Mn-site $(N_{\rm eff})$ defined as, $N_{\rm eff}=\frac{2m_0V}{\pi e^2}\int_{\omega_1}^{\omega_2}\omega Im[\epsilon(\omega)\mu(\omega)]d\omega,$ where $m_0$ is the free electron mass, $e$ the elementary charge, and $V$ the unit-cell volume. We chose $\omega_1=1.8$ meV and $\omega_2=6.2$ meV to characterize the spin excitation. For comparison, we also plot $N_{\rm eff}$ for $E^\omega \parallel b$ and $H^\omega \parallel a$ (triangles) and for $E^\omega \parallel c$ and $H^\omega \parallel a$ (circles). By the above definition, $N_{\rm eff}$ includes the contribution of $\mu$ component driven by $H^\omega$, which is however very small and crudely estimated to be $0.04\times10^{-4}$ and $0.05\times10^{-4}$ for $H^\omega\parallel c$ and $H^\omega\parallel a$, respectively, at 10 K. These values are obtained from $Im[\mu]$ spectrum, which is extracted from $Im[\epsilon\mu]=Re[\epsilon]Im[\mu]+Im[\epsilon]Re[\mu]$ by assuming two Lorentz oscillators for $\epsilon$ and $\mu$. $N_{\rm eff}$ for $E^\omega \parallel a$ increases from well above $T_{\rm N}^{\rm Mn}$ and sharply enhances around $T_{\rm N}^{\rm Mn}$. This is in contrast to $N_{\rm eff}$ for $E^\omega \parallel b$ and $E^\omega \parallel c$, both of which are nearly $T$-independent but turn to increase below $T_{\rm N}^{\rm Mn}$, as presented on a magnified scale in the inset of Fig. \ref{fig3}. Such enhancement mainly arise from the evolution of $\mu$ component by $H^\omega\parallel a$, as mentioned above. As further decreasing $T$, $N_{\rm eff}$ for $E^\omega \parallel a$ reaches the maximum in the FE $ab$ spiral spin phase. Namely, the anisotropy ratio of $N_{\rm eff}$ polarized along the $a$-axis to other axes increases from $\sim3$ at 250 K to $\sim16$ around $T_{\rm c}$. Note that there is discerned the appreciable $a$-polarized absorption above $T_{\rm N}^{\rm Mn}$, manifesting the persistent spin fluctuation. Finally, $N_{\rm eff}$ is considerably suppressed in the $A$-type AFM phase below $T_{\rm N}^{\rm WFM}=15$ K.

In the $ab$ spiral spin phase, the rotation mode of the spiral spin plane is expected to be electric-dipole active for $E^\omega \parallel c$,\cite{HKatsura2} as recently argued to explain the inelastic neutron scattering spectrum in the $ab$ spiral spin phase of TbMnO$_3$ induced by $H$.\cite{DSenff2} However, no apparent peak structure being comparable to the case for $E^\omega\parallel a$ is found in $\epsilon\mu$ spectra for $E^\omega \parallel c$, as we confirmed by measurements using the $ac$- and $bc$-surface crystal plates shown in Figs. \ref{fig2}(d) and \ref{fig2}(c), respectively. Therefore, the present results can firmly exclude the possibility that the spin excitation for $E^\omega\parallel a$ comes from the rotation of the spiral spin plane. Combined with the results of the negligible effect of $H$ on spin excitations in DyMnO$_3$ for both $E^\omega\parallel a$ and $E^\omega\parallel c$ (ref. 11), we can conclude that there is a unique selection rule along the $a$-axis for the strongly infrared active spin excitations in a family of $R$MnO$_3$.

As the possible candidate for the strongly dipole-active spin excitation, the two-magnon process was discussed in the study of DyMnO$_3$ by considering the large mutual fluctuation within the $ac$ plane, inherent to the AFM interaction along the $c$-axis.\cite{NKida} In the present work, the spin excitation for $E^\omega \parallel a$ also broadly distributes over the measured energy range (1--10 meV) [Fig. \ref{fig2}(a)], while the AFMR of Mn ions or $k=0$ magnon driven by $H^\omega \parallel a$ is only identified around 2 meV in the narrow energy range (1--4 meV) [Figs. \ref{fig2}(d) and \ref{fig2}(e)]. In the $ab$ spiral spin phase of TbMnO$_3$ in $H=120$ kOe, a recent neutron scattering study has revealed three branches of the magnon bands around 0.44 meV, 2.2 meV, and 3.06 meV at $k=0$.\cite{DSenff2} Among them, the second mode ($\sim$ 2.2 meV) can be assigned to the $k=0$ AFMR we identified around 2 meV for the case of $H^\omega \parallel a$. Apparently, the broad spectral shape for $E^\omega\parallel a$ cannot be explained by assuming the $k=0$ magnon (Goldstone mode like) alone within the $ab$ plane. Therefore, it is reasonable to consider that it reflects the density-of-states of the magnon band, whose lower and upper edges locate at about 1 and 8 meV, respectively;\cite{DSenff} the two-magnon process is thus most probable for the electrically driven spin excitation continuum in the measured photon energy.\cite{TMoriya} In this picture, the low- and high-lying peak structures can be assigned to van Hove singularities of the two-magnon excitation continuum band. In $k\neq0$, the spin fluctuation still persists well above $T_{\rm N}^{\rm Mn}$, producing the finite $N_{\rm eff}$, as we indeed observed here. In the $A$-type AFM phase, the in-plane two-magnon process is prohibited for the $ab$ plane ferromagnetic spins;\cite{TMoriya} this is also consistent with the observed steep reduction of $N_{\rm eff}$ below 15 K.

In summary, a rich variety of spin excitations (1--10 meV) in thermally induced phases of Gd$_{0.7}$Tb$_{0.3}$MnO$_3$ have been revealed by THz time-domain spectroscopy. We firmly identify the broad electric-dipole active spin excitation only for $E^\omega \parallel a$ in the $ab$ spiral spin phase, clearly excluding the possibility of the rotation mode of the spiral spin plane (one magnon) as the origin of this excitation. The observed unique characteristics of the spin excitations can be explained by the electric-dipole active two-magnon process.

\section*{Acknowledgements}

We thank S. Miyahara and N. Furukawa for fruitful discussions. This work was in part supported by Grant-In-Aids for Scientific Research (16760035 and 17340104) from the Ministry of Education, Culture, Sports, Science and Technology (MEXT), Japan.

\newpage
\begin{figure}

\caption{(Color online) Phase diagrams of (a) TbMnO$_3$ and (b) Gd$_{0.7}$Tb$_{0.3}$MnO$_3$, as reproduced from the data in refs. 4, 5, and 16, with schematic illustrations of the spin structures in $bc$ and $ab$ spiral spin phases and $A$-type antiferromagnetic (AFM) phase. In TbMnO$_3$, $ab$ spiral spin phase ($P_{\rm s}\parallel a)$ is induced in relatively high $H$, which smoothly connects to the $ab$ spiral spin phase of Gd$_{0.7}$Tb$_{0.3}$MnO$_3$ in zero $H$.}
\label{fig1}

\caption{(Color online) Temperature variations of the low-energy electrodynamics of Gd$_{0.7}$Tb$_{0.3}$MnO$_3$ in a variety of thermally induced phases. Upper and lower panels show the real and imaginary parts of $\epsilon\mu$ spectra (closed circles) when $E^\omega$ and $H^\omega$ were set parallel to the crystallographic axes of $ac$-, $ab$-, and $bc$-surface crystal plates. For the case of $E^\omega\parallel a$, the $\epsilon\mu$ spectra in the paraelectric collinear spin phase can be fitted with a Debye relaxation model, and those in the ferroelectric $ab$ spiral spin phase with two Lorentz oscillators for $\epsilon$, as shown by solid lines. In the $A$-type antiferromagnetic phase, the $\mu$ component becomes active and the $\epsilon\mu$ spectrum can be fitted with two Lorentz oscillators for $\epsilon$ and $\mu$ (solid lines).}
\label{fig2}

\caption{(Color online) Temperature dependence of the spectral weight $N_{\rm eff}$, defined in the text, of Gd$_{0.7}$Tb$_{0.3}$MnO$_3$ for $E^\omega\parallel a$ and $H^\omega\parallel c$ (squares), for $E^\omega\parallel b$ and $H^\omega\parallel a$ (triangles), and for $E^\omega\parallel c$ and $H^\omega\parallel a$ (circles). The solid lines are merely the guide to the eyes. The inset shows the expanded view below 45 K. The data indicated by circles and triangles in the inset are multiplied by a factor of 10 for comparison. }
\label{fig3}
\end{figure}

\clearpage
\begin{figure}[bt]
\begin{center}
\includegraphics[width=0.8\textwidth]{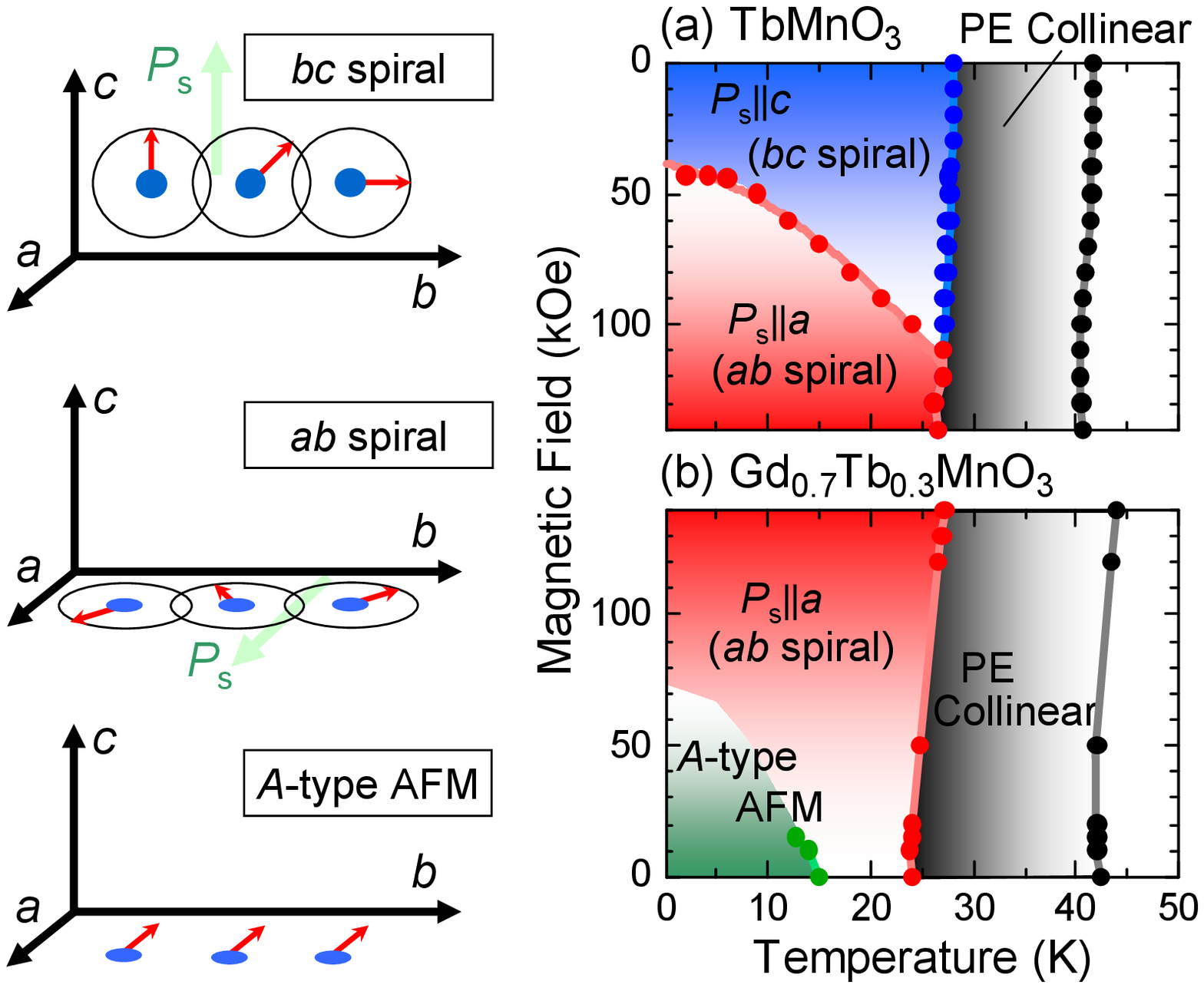}
\end{center}
\end{figure}

\clearpage
\begin{figure}[bt]
\begin{center}
\includegraphics[width=0.8\textwidth]{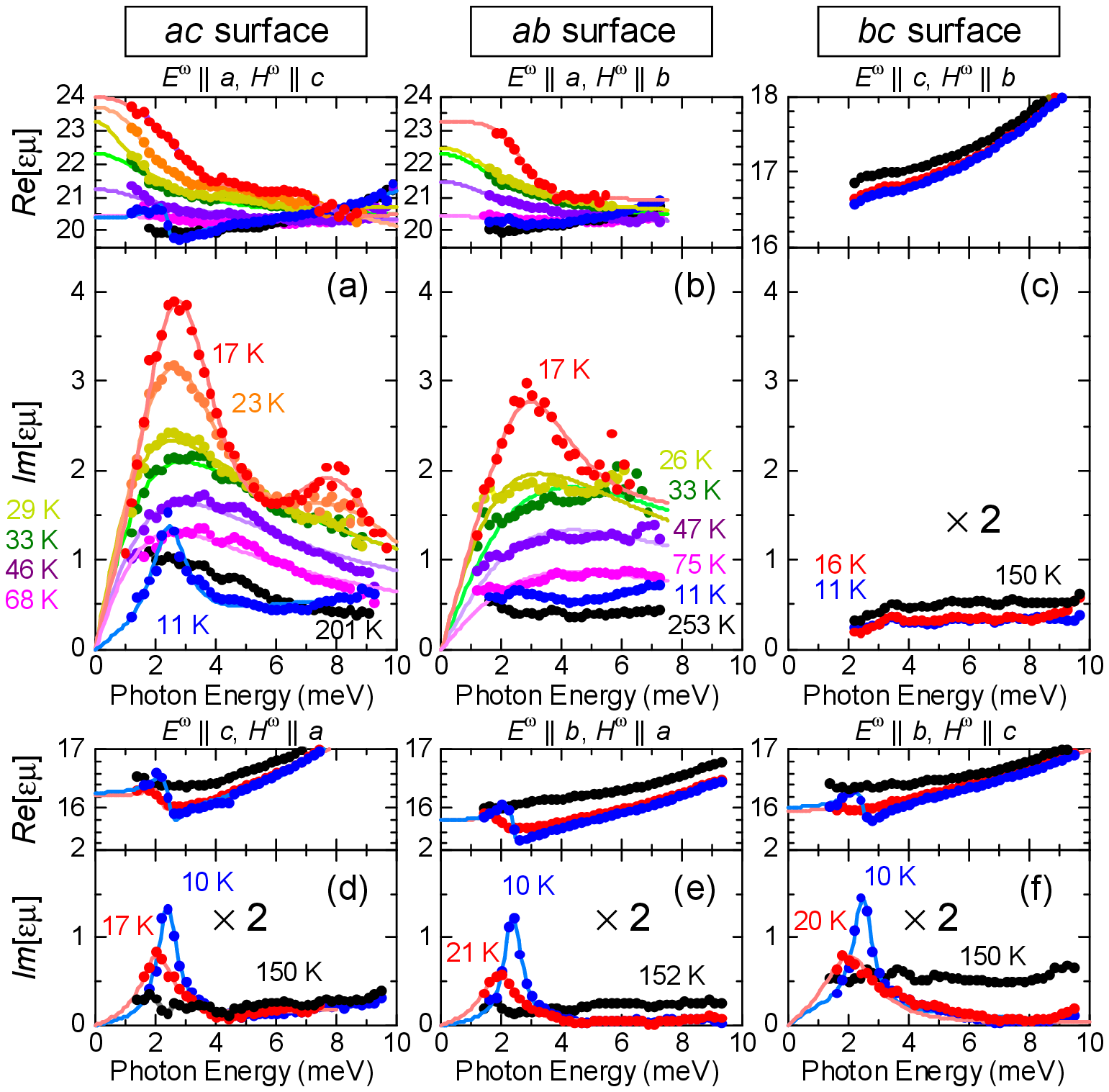}
\end{center}
\end{figure}

\clearpage
\begin{figure}[bt]
\begin{center}
\includegraphics[width=0.8\textwidth]{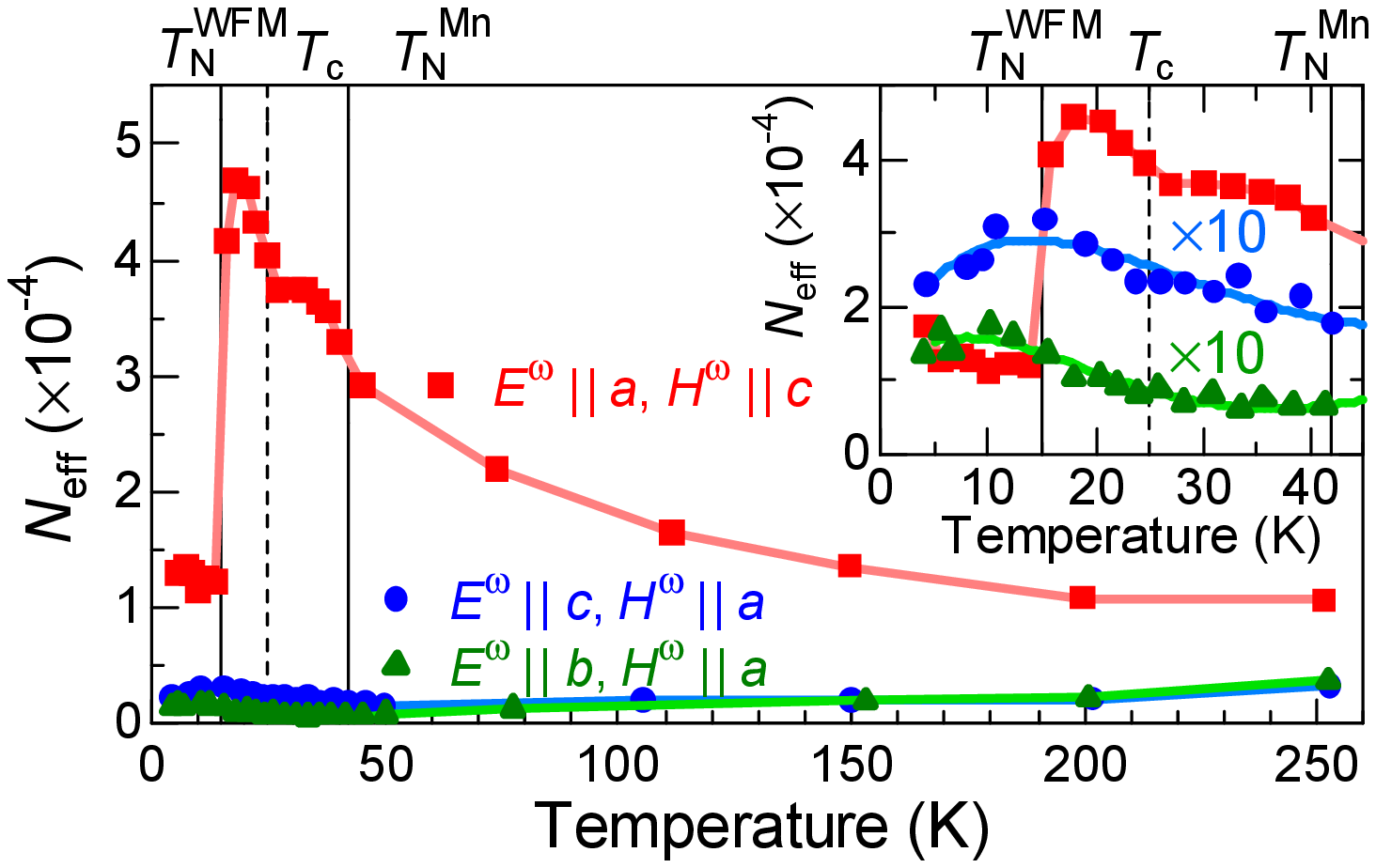}
\end{center}
\end{figure}

\end{document}